\documentclass[aps,prl,twocolumn]{revtex4}
\usepackage{graphicx}
\usepackage{bm}
\usepackage{dsfont}

\newcommand{\be}{\begin{equation}}
\newcommand{\ee}{\end{equation}}
\newcommand{\ben}{\begin{eqnarray}}
\newcommand{\een}{\end{eqnarray}}
\newcommand{\bes}{\begin{subequations}}
\newcommand{\ees}{\end{subequations}}

\newcommand{\bra}[1]{\langle{#1}|}
\newcommand{\ket}[1]{|{#1}\rangle}

\begin{document}
\title{Non-Markovianity assisted Steady State Entanglement}

\author{Susana F. Huelga$^1$, \'Angel Rivas$^{1,2}$ and Martin B. Plenio$^{1,3}$}
\affiliation{$^1$ Institut f\"{u}r Theoretische Physik, Albert-Einstein-Allee 11, Universit\"{a}t Ulm, D-89069 Ulm, Germany}
\affiliation{$^2$ Departamento de F\'{\i}sica Te\'orica I, Universidad Complutense, 28040 Madrid, Spain}
\affiliation{$^3
$ QOLS, The Blackett Laboratory, Prince Consort Road, Imperial College, London, SW7 2BW, UK}

\begin{abstract}
We analyze the steady state entanglement generated in a coherently coupled dimer system subject to dephasing noise as a function of the degree of
Markovianity of the evolution. By keeping fixed the noise strength while varying the underlying dynamics, we demonstrate that non-Markovianity
is an essential, quantifiable resource that may support
the formation of steady state entanglement whereas purely Markovian dynamics
governed by Lindblad master equations lead to separable steady states. This
result illustrates possible mechanisms leading to long lived entanglement in
purely decohering, possibly local, environments. We present a feasible experimental demonstration of
this noise assisted phenomenon using a system of trapped
ions.
\end{abstract}

\maketitle

The generation and ultimate persistence of quantum
entanglement is normally thought to be correlated with a
high degree of system isolation, while the presence of a
surrounding environment tends to decohere the quantum
system thus driving it towards a classically correlated state.
However, driven systems, generally out of equilibrium,
have been shown to tend towards steady states where
quantum correlations are nonvanishing \cite{prl2002}.
A variety of entanglement preserving mechanisms have been put forward,
showing how the presence of environmental noise can be instrumental in keeping the
system entangled in the steady state \cite{prl2002,briegel,brandes,prl2007,rumano,zueco,morigi}.
In the framework of a Markovian dynamics, the presence of local pure dephasing
is normally detrimental for entanglement preservation \cite{ejpb}. The situation
though is very different when the dephasing noise is non-Markovian.
We show that the presence of steady
state entanglement can be unambiguously linked to an increasing degree of non-Markovianity, even when the
environments are acting locally. This result is of particular interest in the
light of identifying mechanisms that assist entanglement preservation in condensed
matter and biomolecular systems, where non-Markovian dephasing is a dominant noise
source \cite{lofranco,caruso,tanimura,cpsun,aspuru}.\\
\begin{figure}[t]
\centerline{\includegraphics[width=.45\textwidth]{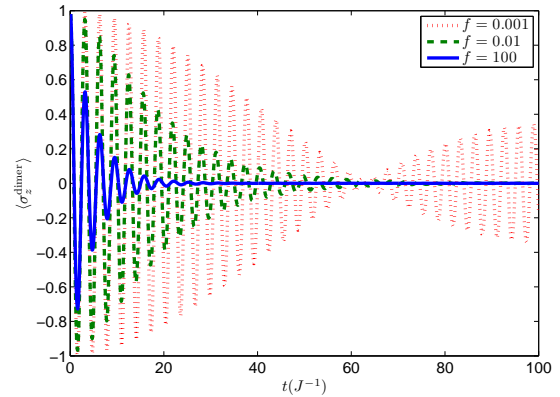}}
\caption{Beating pattern in the dimer population inversion for different values of $f$ (See main text for details).  A Markovian environment ($f \gg 1$) will simply wash out the discreteness of the vibrational modes while the presence of beating in the dimer signal is a signature of the persistence of a coherent interaction with a localized vibrational (damped) mode.}\label{nmm}
\end{figure}
\paragraph{The system --}
We consider a dimer system coherently coupled via an exchange interaction of strength $J$, so
that the system Hamiltonian reads
$H_s = \sum_{j=1}^2 \omega_j \sigma_j^{+}\sigma_j^{-} + J (\sigma_1^{-} \sigma_{2}^{+}
+ \sigma_1^{+}\sigma_{2}^{-})$ $(\hbar =1)$, and subject to the action of an environment
that leaves the populations unaffected but tends to randomize the phases of superposition
states (the so-called pure dephasing or transverse decoherence). We will model this
situation by subjecting each element $j=1,2$ of the dimer to the action of a {\em localized}
mode, being the system-mode interaction governed by a Hamiltonian of the form $H_{s-m}= g_j \sigma_z^j (a_j + a_j^{\dagger})$, where
$a_j (a_j^{\dagger})$ denote the the operators  of annhilitation (creation) of mode excitations.
We will assume that the local modes are damped by a conventional Markovian bath so that the global time
evolution of the dimer and the vibrational modes is described by a Markovian master equation
with a Liouvillian part accounting for the damping of the localized modes at a certain rate
$\kappa$. Note, however, that tracing out the local mode leads to a density matrix for the dimer system that is in general no reproducible
from a purely Markovian evolution for the dimer alone.
We will show that the interplay between the entangling exchange interaction and
the local dephasing will result in an entanglement dynamics that depends crucially on
whether of not the action of the environment can indeed be described by merely subjecting the
dimer to Markovian dephasing.
In the limit where the coherent coupling $g$ is smaller than the decay $\kappa$ we
find that the effective dephasing rate $\gamma_{\mathrm{eff}}$, which quantifies the strength of the system-environment coupling, is proportional to the ratio
$g^2/\kappa$ (see \cite{atach} for the case of cavity QED). When $2g\ll\kappa$ the
dephasing is exponential and the dynamics of the system can be reproduced by a
Lindblad master equation with the effective decay rate $\gamma_{\mathrm{eff}}$.
For $2g<\kappa$, a regime which we denote as weakly non-Markovian, the definition
of an effective dephasing rate $\gamma_{\mathrm{eff}}$ is still reasonable for times
exceeding $\kappa^{-1}$ when the decay is well approximated by an exponential.
In the limit where $g > \kappa$ significant coherent oscillations can occur and
the definition of an effective dephasing rate becomes meaningless.
In this work we will always be operating in the regime where $2g < \kappa$ and
will analyze the different dynamics that the dimer will undergo when subject to
the same effective noise strength but experiences a different underlying dynamics. This will allow us to single out the
precise influence of environmental time-correlations on the persistence/absence of
stationary quantum correlations in the dimer system.

\paragraph{A convenient parametrization--} In order to link the observed
dynamics to a parameter that quantifies in some form how much the environmental action
departs from strict Markovianity, we will introduce an index $f$ such that the
coherent spin-mode coupling $g$ and the mode damping rate $\kappa$ are given
by, respectively: $g=\sqrt{f} g_0$ and $\kappa=f \kappa_0$. In this way, the
noise strength $g^2/\kappa = g_0^2/\kappa_0$ is kept fixed while varying $f$ from
values much larger than $1$ to much smaller than $1$ leads to, respectively,
Markovian and non-Markovian dynamics with, crucially, the same effective local
dephasing rate $\gamma_{\mathrm{eff}}$. In other words, the {\em strength} of
the dephasing noise is kept fixed, but the underlying noisy dynamics is modified,
making the coherent system-oscillator coupling to dominate
over losses (small $f$ domain) or viceversa (limit of large $f$). Note, however, that while the decoherence
rate is kept constant as the parameter f is varied, the
population decay (which depends on $g$ and $\kappa$) does change,
as shown in Fig. 1. The monotonic change in Markovianity
of the noise with the parameter f is shown in Fig. 2, where
an explicit measure of non-Markovianity is considered.
\paragraph{Analytical Results.--}
The effective Hamiltonian for the dimer-local damped
modes system takes the form
\ben
    H &=& \omega_1 \sigma_1^+\sigma_1^- + \omega_2 \sigma_2^+\sigma_2^-
    + J(\sigma_1^+\sigma_2^- + \sigma_1^-\sigma_2^+) \\ \nonumber
    && + (\Omega_1-i\kappa_1)a_1^{\dagger}a_1 + (\Omega_2-i\kappa_2)a_2^{\dagger}a_2\\ \nonumber
    && + g_1 \sigma_1^{(z)}(a_1+a_1^{\dagger})+ g_2 \sigma_2^{(z)}(a_2+a_2^{\dagger}),
    \nonumber
\een
where $\omega_i$ and $\Omega_i$ ($i=1,2$) denote the site and the mode frequencies, respectively, and
\ben
    \dot\rho &=& -iH\rho+i\rho H^{\dagger} + 2\kappa_1a_1\rho a_1^{\dagger} +
    2\kappa_2a_2\rho a_2^{\dagger}.
    \label{hamil}
\een
Motivated by the situations frequently encountered
in biomolecular complexes where external illumination is either weak and/or doubly excited states are strongly suppressed \cite{bio}, we will focus here on situations where the dimer dynamics is confined to the one excitation sector so that the entire dynamics for the two sites takes place
in the subspace spanned by the states $|01\rangle,|10\rangle$.
By introducing the convenient delocalized states
$|u\rangle = \frac{1}{\sqrt{2}}(|01\rangle + |10\rangle)$ and $|d\rangle = \frac{1}{\sqrt{2}}(|01\rangle - |10\rangle)$
and the operators
\ben
    \sigma_{p,m} &=& \frac{1}{\sqrt{2}}(\sigma_1^{(z)} \pm \sigma_2^{(z)}),\\
    a_{p,m} &=& \frac{1}{\sqrt{2}}(a_1 \pm a_2),
\een
noting that $\sigma_p|u\rangle = \sigma_p|d\rangle=0$ while
$\sigma_m|u\rangle=\sqrt{2}|d\rangle$ and $\sigma_m|d\rangle=\sqrt{2}|u\rangle$,
in the subspace spanned by $|u\rangle$ and $|d\rangle$, the Hamiltonian part accounting for the system-mode interaction is
given by
\ben
    H_{s-m} =  \frac{g_-}{2} \sigma_m (a_p + a_p^{\dagger})+ \frac{g_+}{2} \sigma_m (a_m + a_m^{\dagger}),
\een
where $g_{\pm}=g_1 \pm g_2$. Note that the restriction to the one excitation sectors removes the contribution from terms
involving $\sigma_p$, as this operator does not couple
to the states $|u\rangle$ and $|d\rangle$.
In the special case where $\omega_1=\omega_2=\omega, \Omega_1=\Omega_2=\Omega, \kappa_1=\kappa_2=\kappa$
and $g_1=g_2=g$, the total Hamiltonian takes the form
\ben
    H &=& (\omega+J)|u\rangle\langle u| + (\omega-J)|d\rangle\langle d|\\ \nonumber
    && + (\Omega - i\kappa)a_m^{\dagger}a_m
     + \sqrt{2}g(|u\rangle\langle d|+|d\rangle\langle u|)(a_m+a_m^{\dagger}),\nonumber
\een
and the time evolution reads,
\ben
    && \dot\rho = -iH\rho+i\rho H^{\dagger} + 2\kappa a_m\rho a_m^{\dagger}.
\een
An analogous derivation could be done for the case where both subsystems in the dimer couple to a global mode. In this case,
$H_{s-m} = \sigma_m g (a+a^{\dagger})$,
and therefore the dimer would remain decoupled if $g_1=g_2$, while different couplings would yield to a system's dynamics analogous to that of the
effective qubit $(\ket{d},\ket{u})$ discussed above with $g=g_1-g_2$.
\begin{figure}[t]
\centerline{\includegraphics[width=.45\textwidth]{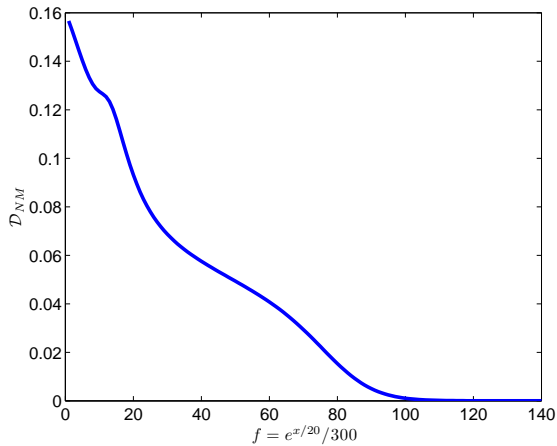}}
\caption{Using the non-Markovianity measure introduced in \cite{nmm} we can rigourously quantify the deviation from strict Markovianity of the dynamical map describing the time evolution of the dimer system.
A finite value of the measure $\mathcal{D}_{NM}$ indicates that the dynamics cannot be accounted for in terms of a master equation of the Lindblad form. Here the values of $f$ range from $f_{min}=0.0035$ to $f_{max}=3.6554$
when the coordinate $x$ varies between 0 and 140. The value of $\mathcal{D}_{NM}$ decreases monotonically with $f$ for the selected parameter regime. For $f$ sufficiently large, the dynamical
map is divisible, $\mathcal{D}_{NM}$ is zero and the dimer's evolution is fully Markovian. }\label{nmmfigure}
\end{figure}

Figure 1 depicts the time evolution of the dimer's population inversion $\sigma_z^{dimer}=\ket{10}\bra{10}-\ket{01}\bra{01}$ for different values
of $f$. For this example the following parameters were considered:
$\omega=0$, $J=1$ (interdimer coupling), $\Omega$=2$J$,
$g_0=J$ and $\kappa_0 = 20 J$. With this, $\gamma_{\rm{eff}}=J/10$.
We truncated each oscillator after Fock layer $2$ as the probability to
have a single excitation in either mode never exceeds 10$\%$ in the parameter
regime that we consider here. Varying $f$ in the range from $10^{-3}$ to
$100$ allows us to move from a situation where $g \ll \kappa$ to the domain
where $g \sim \kappa$, so that the damped mode ranges from being Markovian
to imprinting an element of non-Markovianity to the dynamics, as quantitatively
exemplified when evaluating the degree of non-Markovianity $\mathcal{D}_{NM}$
introduced in \cite{nmm}. This measure provides a necessary and sufficient condition for a given evolution
to depart from strict Markovianity by means of evaluating whether or not the associated
dynamical map $\mathcal{E}_{(t+\epsilon,t)}$ is completely positive (CP) for any $\epsilon$. For that,
it has to be $\left(\mathcal{E}_{(t+\epsilon,t)}\otimes\mathds{1}\right)|\Phi\rangle\langle\Phi|\geq0$,
where $\ket{\Phi}$ is maximally entangled of our open system
and some ancillary system \cite{choi}. Using this condition, one can define a quantitative measure to quantify how much does the evolution departs
from strict Markovianity by evaluating $
\mathcal{I}=\int_0^\infty g(t)dt$, where the function $g$ quantifies how much does the norm of state $\left(\mathcal{E}_{(t+\epsilon,t)}\otimes\mathds{1}\right)|\Phi\rangle\langle\Phi| $ differ from 1 in the limit of $\epsilon\rightarrow 0$ \cite{nmm}.
In Figure 2, we have plotted the normalized measure $\mathcal{D}_{NM}=\frac{\mathcal{I}}{\mathcal{I}+1}$
for a range of values of $f$, showing that the domain of large $f$ leads to
the dimer's dynamics to be fully Markovian.
%
Limiting the harmonic oscillator to just 2 levels, we find for the steady
state population of the state $|d\rangle$ the result
\be
    \rho_{dd}^{\mathrm{ss}}\sim\frac{4g^2+\kappa^2 + (2J+\Omega)^2}{2(4g^2+\kappa^2+4J^2+\Omega^2)}.
\ee
In the limit of sufficiently small $f$, and in the regime where the local mode is quasi-resonant with the dimer eigenstates, so that $\Omega \sim J$,
one finds that the above expression tends to unity. Hence, in this picture,
the system is in weak interaction with a composite environment (local mode+Markovian reservoir) and relaxes towards
its ground state, which in this case is the singlet state. When weakly coupled to a purely Markovian environment and experiencing dephasing at the same rate as before,
the steady state of the dimer system tends towards the maximally mixed state and no entanglement survives in the long term.
The time evolution of the entanglement in the dimer system, as quantified by
the logarithmic negativity \cite{logneg} of the state, is summarized in Figure
3, for an initial factorized condition where there is an excitation in one
site only. Indeed, when $f$ is large, the dynamics of the dimer can be reproduced
by simulating a Markovian master equation with local dephasing rates $\gamma_{\mathrm{eff}}$.
In this case, the steady state of the dimer is maximally mixed, a result that
can be proven using the results presented in section 4 of \cite{ejpb}, following
theorem 5.2 of \cite{spohn}. When $f$ decreases but remaining in the weakly
non-Markovian regime, the steady state approaches the singlet state, which
in this case is the lowest energy eigenstate of the coherent evolution.
Provided that we keep on the weak non-Markovian regime (so that the condition $\kappa_i > g_i$ is satisfied), altering the symmetry of the problem and considering $g_1 \neq g_2$ does not reduce the overlap with the singlet
below $90 \%$ up to ratios $g_2/g_1=2$.
At a finite temperature, the
steady state entanglement is decreased but remains finite. For an average photon number of $n_{th}=0.1$, which corresponds to a temperature of about 100 mK for GHz frequencies, we obtain that the steady state entanglement reaches the value 0.85 for $f=0.01$ (Finite $T$ results are not shown in the Figures).
\begin{figure}[t]
\centerline{\includegraphics[width=.45\textwidth]{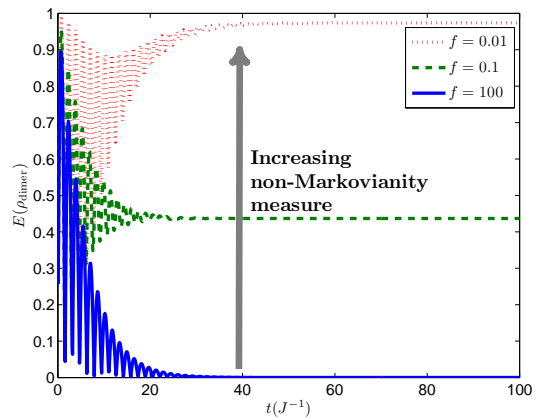}}
\caption{Time evolution of the entanglement content of the dimer system when the index $f$ varies from the range $f\gg1$ (Markovian evolution) to $f\ll1$ (non-Markovian effects). When the evolution is strictly Markovian, the steady state is separable while in the presence of non-Markovianity, the dimer system contains quantum correlations that steadily become close to one e-bit for $f$ sufficiently small.
}\label{enta}
\end{figure}
%

%
\paragraph{Possible experimental realization --}
In this section we outline a possible implementation of the dimer in contact
with non-Markovian environments that was discussed in the previous section.
This realization employs proven elements of ion trap technology and offers, in principle, the
possibility for control of all the relevant parameters in the Hamiltonian
Eq. (\ref{hamil}). Three basic dynamical elements have to be generated. The electronic
degrees of freedom of trapped ions constitute the dimer and need to be coupled to each other,
the motional degrees of freedom play the role of the environment and need to
be coupled to the electronic degrees of freedom via a dispersive interaction
and laser cooling needs to be employed to induce damping of the motional degrees
of freedom which generates a Lorentzian spectral density.

For the implementation of the direct coupling between the two qubits we make
use of a S{\o}rensen-M{\o}lmer gate \cite{SorensenM99,Sackett2000}. To this end one applies laser
fields with two different frequencies so that the two-photon process coupling
$|gg\rangle \leftrightarrow |ee\rangle$ and $|eg\rangle \leftrightarrow |ge\rangle$
is resonant, i.e., $\omega_1^L + \omega_2^L = 2\omega_{eg}$, while neither of the
frequencies are resonant with single excitations of the ions. This can be achieved
by choosing $\omega_1^L  = \omega_{eg} - \delta$ and $\omega_2^L = \omega_{eg} + \delta$,
where $\delta \ll \eta\Omega_R,\nu$. Here $\eta$ denotes the Lamb-Dicke parameter,
$\Omega_R$ the laser Rabi frequency and $\nu$ is the frequency of the center-of-mass mode.
If we prepare the two ions representing the dimer initially in a single excitation
subspace we realize in this fashion the Hamiltonian $H_{\rm dimer} = J_{\mathrm{eff}}(\sigma_1^+\sigma_2^-
+\sigma_2^+\sigma_1^-)$ with an effective exchange coupling $J_{\mathrm{eff}}\sim (\Omega_R \eta)^2/(\nu-\delta)$ \cite{SorensenM99}. The
added advantage of this scheme is the fact that it does not require the
center-of-mass mode to be in its ground state thus increasing the robustness of the scheme.\\
The excitation preserving coupling Hamiltonian $\sigma_z(a+a^{\dagger})$ between an ion
and the motional degrees of freedom can be achieved in several ways. One
approach \cite{PorrasMvD+08} subjects the ion to a far off-resonant standing wave,
which creates the state dependent potential $V(z) = V_0 \cos^2 (k{\hat z} +
\frac{\pi}{4})\sigma_z$. Here $k$ is the wave-vector of the standing wave
lasers. The operator ${\hat z}$ is readily expressed in terms of phonon
operators, $z = \sum_n {\cal M}_n\sqrt{\frac{1}{2m\omega_n}} (a_n + a_n^{\dagger})$,
where $m$ is the ion mass, and ${\cal M}_n$ is the amplitude of each vibrational
mode $n$ at the ion. Expanding this potential in the small parameter $kz\ll 1$
we find that the leading order contribution is linear in $z$ with the second
order contribution canceling. Hence we obtain a Hamiltonian $H_{sb} = \sigma_z
\sum_n {\cal M}_n\sqrt{\frac{1}{2m\omega_n}} (a_n + a_n^{\dagger})$. Traveling
wave fields can also be used to generate this type of interaction in a suitably
transformed basis \cite{PorrasMvD+08}. This scheme couples to all the modes of
the ion crystal which may lead to correlations between
the environments acting on separate parts of the dimer. We may select specific
environment modes by choosing light fields that are slightly off-resonant to
specific modes while being more strongly detuned from the remaining modes.\\
Finally our dimer model assumes that the modes coupling to the constituents of the
dimer are damped. Damping of motional degrees of freedom can be achieved, of course, by
applying laser cooling to an auxiliary third ion that couples to all the motional
modes of the system. To be effective, this auxiliary ion should be placed at the
end of the ion chain to ensure coupling to all modes in the ion string.
Combining these three elements yields the dynamical Hamiltonian discussed
in Eq. (\ref{hamil}) which underlies the non-Markovianity driven steady state entanglement
in a dimer system.

\paragraph{Conclusions --} We have demonstrated that, when subject to system-environment interactions of the same strength, the non-Markovian character of the noise can be the crucial property that leads to steady state entanglement where purely Markovian noise would result in the complete destruction of entanglement. A possible experimental verification of this depahsing-assisted phenomenon in ion trap physics which employs only experimentally demonstrated building blocks has been discussed. We expect these studies to contribute towards the
identification of the physical mechanisms that could underpin the persistence of stationary quantum correlations in very noisy environments occurring in natural conditions \cite{lofranco,caruso,tanimura,cpsun,aspuru}. A
key issue in this context would be the evaluation of this
effect for realistic (possibly strongly non-Markovian), multicomponent,
biological systems operating at physiological
temperatures. Numerical methods able to incorporate temperature dependence and the ability to simulate exactly structured spectral densities would be required to fully address this issue and initial steps towards this development have already been presented \cite{prior}.

\paragraph{Acknowledgments --}
We are grateful to Felipe Caycedo-Soler for carefully reading the manuscript. This work was supported by the EU STREP
projects CORNER and PICC, the EU Integrated Project Q-Essence, the project QUITEMAD S2009-ESP-1594 of the Consejer\'{\i}a de Educaci\'{o}n de la Comunidad de Madrid, MICINN FIS2009-10061 and by the Alexander
von Humboldt Foundation.

\end{document}